\documentstyle[12pt]{article}
\newcommand{\leqsim}{\,\raisebox{-0.6ex}{$\buildrel < \over \sim$}\,}
\newcommand{\geqsim}{\,\raisebox{-0.6ex}{$\buildrel > \over \sim$}\,}
\def\c{{\rm c}}
\def\con{{\rm contrived}}
\def\df{{\rm d}}
\def\e{{\rm e}}
\def\etal{{\it et al}\,}
\def\hc{{\rm h.c.}}
\def\ibid{{\it ibid}\,}
\def\O{{\cal O}}
\def\P{{\rm Pl}}
\def\rad{{\rm rad}}
\def\s{{\rm s}}
\def\term{{\rm term}}
\def\walls{{\rm walls}}
\def\gev{\,{\rm GeV}}
\def\mev{\,{\rm MeV}}
\def\sec{\,{\rm sec}}

\begin{document}
\thispagestyle{empty}
\begin{flushright}
{\tt OUTP-95-22P\\ RAL-TR-95-019\\ June 1995\\ hep-ph/9506359}
\end{flushright}
\vspace{5mm}
\begin{center}
{\Large \bf On the Cosmological Domain Wall Problem \\
            for the Minimally Extended \\ Supersymmetric Standard Model}\\
\vspace{15mm} {\large S.A.~Abel$^{a}$, S.~Sarkar$^{b}$\footnote{PPARC
 Advanced Fellow} and P.~L.~White$^{b}$}\\
\vspace{1cm}
{\it $^{a}$Rutherford Appleton Laboratory \\
            Chilton, Didcot OX11 0QX, U.K.\\
\vspace{5mm}
$^{b}$Theoretical Physics, University of Oxford \\
       1 Keble Road, Oxford OX1 3NP, U.K.}  \end{center} \vspace{2cm}
\begin{abstract}
\noindent
We study the cosmology of the Supersymmetric Standard Model augmented
by a gauge singlet to solve the $\mu$-problem and describe the
evolution of the domain walls which are created during electroweak
symmetry breaking due to the discrete $Z_{3}$ symmetry in this
model. The usual assumption, that non-renormalizable terms induced by
gravity (which explicitly break this symmetry) may cause the walls to
collapse on a cosmologically safe timescale, is reconsidered. Such
terms are constrained by considerations of primordial nucleosynthesis,
and also by the fact that by not respecting the $Z_{3}$ symmetry they
induce divergences which destabilise the hierarchy and reintroduce the
$\mu$--problem. We find that, even when the K\"ahler potential is
`non-minimal' (i.e.  when the hidden sector couples directly to the
visible) the model is either ruled out cosmologically or suffers from
a naturalness problem.
\end{abstract}

\pagestyle{plain}
\newpage
\section{Introduction}

The purpose of introducing (softly broken) supersymmetry into the
Standard Model is to bring under control the quadratic divergences
associated with a fundamental Higgs boson and make it `natural' for
its mass to be at the electroweak scale~\cite{mssm}. Yet the minimal
supersymmetric Standard Model (MSSM) has its own naturalness
problem. Its Lagrangian contains a term $\mu~H_{1}H_{2}$ mixing the
two Higgs doublets which are now required to give masses separately to
the up- and down- type quarks. For successful phenomenology $\mu$
should also be of order the electroweak scale but this must now be set
by hand --- the `$\mu$--problem'~\cite{muprob,gm}. To address this
problem, the next-to-minimal supersymmetric Standard Model
(NMSSM)~\cite{nmssm} contains an additional singlet Higgs superfield
$N$. By invoking a $Z_{3}$ symmetry under which every chiral
superfield $\Phi$ transforms as $\Phi\to\e^{2\pi{i}/3}\Phi$, the
allowed terms in the superpotential are now
$\lambda{N}H_{1}H_{2}-\frac{k}{3}N^3$ (in addition to the usual Yukawa
terms generating fermion masses) while the Higgs part of the soft
supersymmetry breaking potential is extended by the inclusion of two
additional trilinear soft terms $A_{\lambda}$ and $A_{k}$ to
\begin{eqnarray}
 V_{\rm soft}^{\rm Higgs} &=
  &- \lambda A_{\lambda}(NH_{1}H_{2} + \hc)
  - \frac{k}{3} A_k (N^3 + \hc) \nonumber\\
  &&+ m^2_{H_{1}} \vert H_{1} \vert^2
  + m^2_{H_{2}} \vert H_{2} \vert^2
  + m^2_{N} \vert N \vert^2,
\end{eqnarray}
where $H_{1}H_{2}=H_{1}^0H_{2}^0-H^-H^{+}$. The $\mu$--term can now be
simply set to zero by invoking the $Z_{3}$ symmetry. An effective
$\mu$--term of the form $\lambda\langle{N}\rangle$ will still be
generated during SU(2)$\times$U(1) breaking but it is straightforward
to arrange that $\langle{N}\rangle$ is of order a soft supersymmetry
breaking mass. Apart from solving the `$\mu$--problem' the NMSSM also
has interesting implications for supersymmetric
phenomenology~\cite{phen} and dark matter~\cite{dm}.

Unfortunately, the NMSSM runs into a cosmological difficulty. The
$Z_{3}$ of the model is broken during the phase transition associated
with electroweak symmetry breaking in the early universe. Due to the
existence of causal horizons in an evolving universe, such
spontaneously broken discrete symmetries lead to the formation of
domains of different degenerate vacua separated by domain walls
\cite{zko,kib}. These have a surface energy density $\sigma\sim\nu^3$
where $\nu$ is a typical vacuum expectation value (vev) of the fields,
here the electroweak scale of $\O(10^2)\gev$. Such walls would come to
dominate the energy density of the universe and create unacceptably
large anisotropies in the cosmic microwave background radiation unless
their energy scale is less than a few MeV~\cite{revs}. Therefore
cosmology requires the $Z_{3}$ walls to disappear well before the
present era. Following the original suggestion by Zel'dovich
\etal\cite{zko}, this may be achieved by breaking the degeneracy of
the vacua, eventually leading to the dominance of the true
vacuum. This happens when the pressure, i.e. the difference in energy
density between the distinct vacua, begins to exceed the tension
$\sigma/R$, where $\sigma$ is the surface energy density of the walls
and $R$ the scale of their curvature. When $R$ becomes large enough
for the pressure term to dominate, the domain corresponding to the
true vacuum begins to expand into the domains of false vacuum and
eventually fills all of space. It was recently argued~\cite{grav} that
gravitational interactions at the Planck scale $M_{\P}$ would
explicitly violate any discrete symmetry, causing just such a
non-degeneracy in the minima of $\O(\nu^5/M_{\P})$ where $\nu$ is a
generic vev (of $\O(M_W)$ in our example). In fact, this suggestion
had been applied already to the NMSSM in the context of string
theories~\cite{eenoqz}. Thus there would appear to be a natural
solution to the cosmological domain wall problem for the NMSSM.

In this paper we study whether this solution is indeed viable. In the
following section we derive the structure of the walls, and show that
the surface energy is approximately $M_W^3$ as expected on dimensional
grounds. We go on to describe the evolution of the walls under the
influence of the tension, the pressure due to the small explicit
$Z_{3}$ breaking and the friction due to particle reflections. In
particular we demonstrate that wall domination of the energy density
of the universe is avoided if the gravitationally induced terms are of
order six or less. This is not however the tightest constraint on the
domain walls; by applying constraints based on primordial
nucleosynthesis we show that the magnitude of $Z_{3}$ breaking must be
$\geqsim10^{-7}\sigma{M_{W}^2}/M_{\P}$, in order to make the walls
disappear before the nucleosynthesis era begining at
$T\sim1\mev$. Thus only operators of dimension five (suppressed by at
most one power of the Planck mass) are permitted. This reduces to
three the number of possible $Z_{3}$ breaking terms which are allowed
in the superpotential or K\"ahler potential and which induce
dimension-5 operators in the effective potential. By inspection we
find that the existence of one or more of these operators implies that
there is no symmetry (discrete, global, gauged, $R$-symmetry or gauged
$R$-symmetry) under which the low-energy singlet can be charged. This
implies that there {\em cannot} be any explanation for the absence of
three allowed low energy operators which include the $\mu$--term
itself as well as quadratic and linear terms in $N$. Thus our first
conclusion is that not only does the NMSSM not solve the
$\mu$--problem, it actually makes things worse by introducing
additional operators and by disallowing any symmetry which would
forbid them.

We then go on to consider the fact that the singlet which appears in
the NMSSM may introduce destabilising divergences~\cite{destab}.
Essentially the problem is that by introducing non-renormalizable
terms together with soft supersymmetry breaking, we may produce
corrections to the potential which are quadratically divergent and
thus proportional to powers of the cut-off $\Lambda$ in the effective
supergravity theory. Since the natural scale for this cut-off is
$M_{\P}$, these can in principle destabilise the hierarchy, forcing
the singlet vev and hence the scale of electroweak breaking to become
very large (at least of order $\sqrt{M_{W}M_{\P}}$). By examining the
possible $Z_{3}$ breaking terms, we demonstrate that the removal of
domain walls by this mechanism indeed destabilises the hierarchy. We
conclude that the two constraints, viz. stability of the hierarchy and
domain walls, cannot be simultaneously satisfied by any
gravitationally suppressed operators which one can add to the
Lagrangian.

We consider alternative ways for dealing with the domain walls.  One
possible solution is to reintroduce the $\mu$ term in the
superpotential in such a way as to avoid the introduction of the
dangerous non-renormalisable operators. If we drop the assumption of
minimality in the K\"ahler potential by allowing certain couplings of
the hidden sector fields to the visible sector (as in Ref.\cite{gm})
we can retain $Z_{3}$ symmetry in the full theory, but break it
spontaneously when supersymmetry is broken. In this way the hierarchy
is not destabilised by tadpole diagrams. However the naturalness
problem cannot be solved even for these more general models.

Finally we consider how gauge singlets may be accommodated in
supersymmetry, without invoking these problems. There appear to be
only a few possibilities, none of which yields a phenomenology bearing
any resemblance to the NMSSM.

\section{Domain Walls in the NMSSM}

When a discrete symmetry is spontaneously broken as the universe
expands and cools, `domains' of the different degenerate vacua form
separated by narrow regions of higher potential called `domain
walls'~\cite{revs}. The structure of these walls may be determined by
finding time-independent solutions to the classical field equations
after imposing the boundary conditions that at the endpoints the
fields should be in distinct vacuum configurations. This has been done
using numerical methods for the NMSSM potential~\cite{aw} and we
reiterate the essential features of the $Z_{3}$ walls. As might be
expected from dimensional arguments and by analogy with the
analytically soluble case of a single real scalar field in a $Z_{2}$
symmetric potential~\cite{revs}, the thickness and energy density of
the walls are of order $\nu^{-1}$ and $\nu^3$ respectively, where
$\nu$ is a typical vacuum expectation value. For naturalness reasons
one would tend to assume that all three vacuum expectation values are
of the same order; however, it is also possible that the singlet vev,
$x$, is much larger than the usual
$\nu=\sqrt{\nu_{1}^2+\nu_{2}^2}=174\gev$. This is in fact quite likely
in the light of recent analyses where unification of soft terms and
gauge couplings is imposed at the GUT scale; the only viable scenarios
are then found to have $x/\nu\geqsim10$ with especially large values
when the Higgs sector Yukawa couplings are very small~\cite{gut}. In
such cases, we would expect the wall to have a much higher surface
energy $\sigma$; indeed we find that this is well approximated by
\begin{equation}
 \sigma \simeq 5\times10^7 \gev^3 \left(\frac{k}{0.1}\right)
                \left(\frac{x}{5\nu}\right)^3,
\end{equation}
when $x$ is at least a few times larger than $\nu$. This formula is
accurate to about a factor of 2 in practice and is very good for large
$x$, relative to both the trilinear soft terms and to $\nu$.

Similarly the thickness of walls is given by
\begin{equation}
 \delta \simeq 2\times10^{-2} \gev^{-1} \left(\frac{k}{0.1}\right)^{-1}
                \left(\frac{x}{5\nu}\right)^{-1}
\end{equation}
which again is most accurate when $x\gg\nu$ and
$x\gg{A_{k},A_{\lambda}}$. We show an example of a wall with large
$x$ in Figure~1. In comparison to the cases shown in Ref.\cite{aw}, we
see that the wall is thinner and the surface energy higher, as expected.

We note that if both $A_k$ and $A_{\lambda}$ are zero, then the $Z_3$
symmetry of the scalar potential becomes a $U(1)$ symmetry, and so the
wall energy falls to zero and its width becomes infinite; in this limit
however we have an axion problem. We find that if $A_k$ or $A_\lambda$
are greater than a few GeV then the wall energy is insensiteve to their
exact values.

Immediately after the electroweak phase transition the universe is
filled with equal volumes of the three degenerate phases. These are
correlated on a length scale which depend on the nature of the phase
transition, varying from $\xi\sim~T_{\c}^{-1}$ for a second-order
transition to $\xi\sim~H^{-1}$ for a strongly first-order
transition~\cite{revs,hodges}. Since the probability for each vacuum
(0.333) is just above the percolation threshold (which for continuum
percolation theories is found to be $0.295$~\cite{percol}), the
universe is then filled with highly convoluted, infinite regions
separated by stable domain walls of typical curvature scale $\xi$,
which rapidly grows to the size of the horizon.

Let us now turn to the dynamics of cosmological networks of such
walls.  As discussed in Ref.\cite{zko}, the most important forces
acting on the walls are surface tension, friction and pressure. The
equation of motion for a quasi-spherical piece of wall moving with
velocity $\dot{R}$ (with $\gamma\equiv~1/\sqrt{1-\dot{R}^2}$) and
having local radius of curvature $R$, is
\begin{equation}
\label{eqmo}
 \frac{\df^2 R}{\df t^2}= -\frac{2}{R \gamma^2}
                     -\frac{\langle n p v \rangle}{\sigma \gamma^3}
                     -\frac{\varepsilon}{\sigma\gamma^3}\ .
\end{equation}

The first term on the RHS reflects the fact that it is energetically
favourable for the wall network to reduce its surface area through
surface tension, and hence small domains will collapse, irregularities
in the surfaces will straighten out, and the correlation length will
increase. This term expresses just the conservation of
energy in the absence of pressure and friction.

The second term on the RHS corresponds to friction arising due to the
interactions of the wall network with the thermal plasma. As particles
reflect off the walls, they exert a force given by the thermally
averaged momentum transfer $\langle{n}{p}{v}\rangle$, where $n$ is the
particle density, $v$ the particle velocity relative to the wall, and
$p$ the momentum perpendicular to the wall. (Actually the friction is
$\propto{v}$ only when $v\ll{c}$.) Friction is clearly important at
times very close to the electroweak phase transition if the top quark
and gauge boson fields are still in equilibrium in the plasma. At
later times, when the number density of these particles is
exponentially suppressed, the main source of friction is the
interaction of the walls with lighter fermions in the plasma. The
constant difference in phase in the mass terms on either side of the
wall (i.e. $\pi/3$ or $2\pi/3$) does not by itself cause any
reflection but rather just a phase shift in the fermion masses (as can
be checked by equating transmission and reflection coefficients at the
wall). In order to estimate the reflection coefficient, it is useful
to describe the space dependent mass by the inverted bell-shaped
function
\begin{equation}
 m^2 (x_{\perp}) = m^2
                   - a^2 \frac{\lambda(\lambda-1)}{\cosh^2 a x_{\perp}}\ ,
\end{equation}
where $x_{\perp}$ is the perpendicular distance from the wall, and $m$
is the mass given to the reflecting particle by the Higgs fields which
comprise the domain wall of width $a^{-1}$. The task of finding the
reflection coefficient (using the Klein-Gordon equation) then reduces
to a known problem, the modified P\"oschl-Teller potential, which can
be solved analytically (see for example Ref.\cite{flugge}). We take
the depth of the well to be $m^2$ and the width $a^{-1}\sim
M_{W}$. The depth parameter is $\lambda=(1+m^2/M_{W}^2)$ and the
reflection coefficient is then found to be
\begin{equation}
 |R|^2 = \frac{\pi^2 m^4}{\pi^2 m^4 + M_{W}^4 \sinh^2 \pi p a}
  \approx \frac{m^4}{p^2 M_{W}^2}\ ,
\end{equation}
where we have taken $M_{W}\gg~p~\gg~m$ as is appropriate once the gauge
bosons and top quark have fallen out of equilibrium. (There is a
region at low energy $|p|<m^2/M_{W}$ in which the particles experience
total reflection~\cite{revs}. However this contribution is
insignificant here, being suppressed by many powers of $m^2/M_{W}$.)
Clearly particles which are heavy, especially the bottom quark, will
be more important here. We can estimate the friction by considering a
particle of mass $m$, when the wall velocity through the plasma, $u$,
is small. Then
\begin{equation}
 \langle n p v \rangle = g \int \frac{\df^3 p}{(2\pi)^3}
  \frac{m^4}{p^2 M_{W}^2} \frac{p_{\perp}^2}{E}
  \left[\frac{1}{\exp(\gamma E + \gamma u p_{\perp})/T)+1}
  - \frac{1}{\exp(\gamma E - \gamma u p_{\perp})/T)+1}\right]\ ,
\end{equation}
where $T$ is the temperature of the plasma, and $g$ is the number
of degrees of freedom of the reflecting particles.
Expanding this function in $u$ keeping the leading term only
and performing the angular integral, we find
\begin{equation}
\label{function}
 \langle npv \rangle = \frac{g u}{8\pi^2} T^4 \frac{T^2}{M_{W}^2} F(x_m),
\end{equation}
where
\begin{equation}
 F(x_{m}) = \int^\infty_{x_{m}}
  \df x\ x_{m}^4 (x^2-x_{m}^2) \frac{\e^x}{(\e^x+1)^2},
\end{equation}
and we have defined $x\equiv{E}\gamma/T$ and $x_{m}\equiv{m}\gamma/T$.
This integral is very well approximated by
\begin{equation}
 F(x_{m}) = x_{m}^5 \left(0.6 \e^{-x_{m}}\right)^3.
\end{equation}
Summing over all the particle species in the plasma, we find that
\begin{equation}
 \langle n p v \rangle = f(T)\ u\ \frac{T^4}{8\pi^2},
\end{equation}
where $f(T)<5\times10^{-4}$ at all temperatures. We show $f(T)$ in
Figure~2 where, apart from omitting the contribution of the up and
down quarks, we have neglected the possible effect of the quark-hadron
phase transition. In the era when pressure is negligible (i.e. when
the typical curvature scale is small), we can calculate the terminal
wall velocity, $u_{\term}$, and establish {\em a posteriori} that our
approximation of small $u$ to obtain eq.(\ref{function}) is indeed
correct, i.e. friction {\em is} important. Substituting the friction
into eq.(\ref{eqmo}), we find
\begin{equation}
 u_{\term} = \frac{16\pi^2}{f(T)} \left(\frac{\sigma}{T^4R}\right).
\label{friction}
\end{equation}
For typical values of the radius, $R\sim~u_{\term}t$, we
see that friction is important only at temperatures above a few
hundred MeV. We therefore conclude that shortly after the quark-hadron
phase transition the walls move with velocities comparable to the
speed of light and so we may safely neglect friction in what follows.

The last term on the RHS in eq.(\ref{eqmo}) is the pressure
corresponding to the difference $\varepsilon$ in the energy density
between the different vacua. As remarked earlier, this will become
dominant when it exceeds the surface tension, i.e. when
\begin{equation}
 \varepsilon > \frac{\sigma}{R}.
\end{equation}
We show this happening in Figure~3, where we have performed a simple
thin wall simulation of a network of domain walls using techniques
similar to those used in ref.\cite{kawano}, and which we have
discussed in more detail elsewhere~\cite{aw}. In the absence of
friction it is convenient to rescale the parameters with some typical
length scale, $R_{0}$, which we choose to be 1~cm, corresponding
approximately to the curvature scale when pressure becomes dominant if
$\varepsilon\sim~M_{W}^5/M_{\P}$. Thus defining $\rho\equiv{R}/R_{0}$, and
$\tau\equiv~t/R_{0}$, eq.(\ref{eqmo}) becomes
\begin{equation}
\label{eqmo2}
 \frac{d^2 \rho}{d\tau^2} = - \frac{2}{\rho\gamma^2}
                            - \frac{\varepsilon R_{0}}{\gamma^3\sigma}.
\end{equation}
Thus there are only two independent parameters in our simulation,
given by the pressure in each of the two false vacua,
$\varepsilon{R_{0}/\sigma}$. Initially, the walls expand under their
own tension, and the structure develops in the manner discussed in
Refs.\cite{kawano,wallevol}. Eventually pressure dominates as expected
and the entire volume is cleared of walls. This contrasts with the
no-pressure case, where one or two horizon-sized walls always
remain~\cite{aw}. The behaviour for different values of the pressure or
surface tension is identical if the time and length, respectively, are
scaled appropriately.

One might consider the possibility that since frictionless,
pressureless walls expand until there is roughly one wall per horizon
scale~\cite{revs}, domain walls may be accomodated by simply assuming
that our local region of space-time just happens to be empty of them,
i.e. that there is a wall lurking just outside our present
horizon. There are at least two objections to this. Firstly the walls
eventually come to dominate the energy density of the universe,
causing unacceptable `power-law' inflation~\cite{seckel}, unless their
separation is many times greater than the present horizon scale, which
is clearly impossible by causality. Secondly, even such a wall outside
the horizon will have a curvature scale comparable to the present
horizon scale and thus induce unacceptably large anisotropy in the
cosmic microwave background~\cite{gz}.

\section{When Walls Collide}

What value of the pressure (i.e. explicit $Z_{3}$ breaking) is
required to safely remove the walls? The crudest estimate we can make
is simply to insist that the walls are removed before they dominate
over the radiation energy density in the universe in order to avoid
wall driven inflation. Since the walls move at close to the speed of
light below the quark hadron phase transition, their curvature scale
will be roughly the horizon size,
$R\sim{t}\sim~M_{\P}/g_{*}^{1/2}T^2$. Since the energy density
of the walls is
\begin{equation}
 \rho_{\walls} \sim \frac{\sigma}{R}\ ,
\end{equation}
and the radiation energy density is $\rho_{\rad}\sim~g_{*}^{1/2}T^4$,
we see that walls dominate the evolution below a temperature
\begin{equation}
 T_{\star} \sim \left(\frac{\sigma}{g_{*}^{1/2} M_{\P}}\right)^{1/2}.
\end{equation}
To prevent this we require the pressure to have become dominant before
this epoch, i.e.
\begin{equation}
 \varepsilon > \frac{\sigma}{R_{\star}} \sim \frac{\sigma^2}{M_{\P}^2}.
\end{equation}
A pressure of this magnitude would be produced by dimension-6
operators in the potential. However, one should consider further
constraints coming from primordial nucleosynthesis, and we find that
only operators of dimension-5 or less are sufficient to satisfy these.
In fact for weak scale walls the time associated with the temperature
$T_{\star}$ is $t_{\star}\sim{M_{\P}^2/g_{*}^{1/2}
M_{W}^3}\sim10^8\sec$, i.e. long after nucleosynthesis. The entropy
produced when the walls collide (which is by now a major proportion of
the total entropy in the universe) is dumped into all the decay
products of neutral Higgs particles, i.e. Standard Model quarks and
leptons. In order to check whether this violates phenomenological
bounds, we compute the relative energy density released in such
collisions, viz.
\begin{eqnarray}
 \frac{\rho_{\walls}}{n_{\gamma}}
  &\sim& \frac{\sigma}{t n_{\gamma}} \nonumber\\
  &\sim& 7\times10^{-11}\gev \left(\frac{\sigma}{M_{W}^3}\right)
   \left(\frac{t}{\sec}\right)^{1/2},
\end{eqnarray}
where we have taken the number of relativistic degrees of freedom in
the plasma to be $g_{*}=43/4$. Detailed consideration of the effects
of high energy particles on primordial nucleosynthesis and on the
2.73~K Planckian spectrum of the microwave background radiation impose
severe upper limits on this parameter~\cite{cosmobounds}. For the
typical values of $\sigma$ in Figure~1, we find that the walls are
required to disappear before the onset of nucleosynthesis at about
$0.1\sec$, as otherwise the hadrons in the showers triggered by the
decay products would alter the neutron-to-proton ratio, resulting in a
$^4$He mass fraction in excess of the conservative observational upper
bound of $25\%$~\cite{hadro}. This means that in order not to disrupt
primordial nucleosynthesis, we require explicit $Z_{3}$ breaking of
magnitude
\begin{equation}
 \varepsilon \sim \lambda^\prime \sigma M_{W}^2/M_{\P},
\end{equation}
with
\begin{equation}
\label{nucbound}
 \lambda^\prime \geqsim 10^{-7}.
\end{equation}

\section{The Return of the $\mu$ Problem}

Having established that one needs dimension-5, $Z_{3}$ breaking
operators to appear in the effective potential, we can consider ways
in which this can be achieved by adding terms to the K\"ahler
potential or superpotential.  We first assume that these are `minimal'
in the sense that they do not contain couplings between the hidden and
visible sectors (which couple only through gravity). Later we shall
consider the most general non-minimal case.  In all cases we find that
there is a naturalness problem associated with the explicit breaking
of the $Z_{3}$ symmetry.

Let us write down the contributions to the supergravity Lagrangian
which explicitly break $Z_{3}$, and which are invariant under the
NMSSM gauge group. These are
\begin{equation}
\label{ops}
 \lambda^\prime \frac{N^4}{M_{\P}}\ , \qquad
 \lambda^\prime \frac{N^2(H_{1}H_{2})}{M_{\P}}\ , \qquad
 \lambda^\prime \frac{(H_{1}H_{2})^2}{M_{\P}}\ ,
\end{equation}
in the superpotential, and
\begin{equation}
 \alpha_{i} \frac{(N + N^{\dagger})(H_{i} H^{i\dagger})}{M_{\P}}\ ,
 \qquad \beta \left(\frac{N^{\dagger}H_{1}H_{2} + \hc}{M_{\P}}\right)/ ,
\end{equation}
in the K\"ahler potential. As in Ref.\cite{destab}, we can absorb the
last two contributions into the superpotential to $\O(M_{\P}^{-1})$ by
making the redefinitions
\begin{eqnarray}
 H_{i} &\rightarrow & \left(1 - \frac{\alpha_{i} N}{M_{\P}}\right)
  H_{i}\nonumber\\
 N &\rightarrow & N - \frac{\beta (H_{1} H_{2})}{M_{\P}},
\end{eqnarray}
and so we shall consider only the first three contributions in what
follows. Inspecting these, we observe that $N$ must be a singlet under
any additional symmetry in order for any one of these terms to exist
in addition to the terms $N^3$ and $N H_{1}H_{2}$ in the low energy
superpotential. In other words, each of them implies that the
following `unnatural' contribution to the superpotential is
invariant
\begin{equation}
 \delta W_{\hbox{\rm `unnatural'}} =
  \mu^{\prime\prime} N + \mu^\prime N^2 + \mu H_{1} H_{2}.
\end{equation}
Thus not only have we reintroduced the $\mu$--problem, we now have two
additional naturalness problems. Whereas the standard $\mu$--problem
may well be solved at a future date (for example by the mechanism of
Ref.\cite{gm}), we shall see that the naturalness problem which has
reappeared here can have no solution based on an underlying symmetry.

\section{The Return of the Hierarchy Problem}

As if the difficulties above were not bad enough, there is the
possibility of quadratic tadpole divergences which can lead to a
destabilisation of the hierarchy~\cite{destab}. This exacerbates our
problems, since such divergences arise at each order in perturbation
theory, forcing us to {\em re}-fine-tune. These are a potential
problem in any supergravity model with gauge {\em singlets} since the
dangerous diagrams are not excluded by gauge invariance.

These diagrams arise when supersymmetry is spontaneously broken,
because super-Weyl-K\"ahler invariance necessitates that the vev of
the K\"ahler potential become non-trivial. In fact~\cite{destab}
\begin{eqnarray}
 \langle\e^{2 K/3}\rangle &\approx &\left.\e^{2 K/3}\right|
  \left(1 + \theta^2 M_{\s}^2 + \bar{\theta}^2 M_{\s}^2 +
  \theta^2 \bar{\theta}^2 M_{\s}^4\right) \nonumber\\
  \langle\phi\rangle &\approx &\left.\phi\right|
  \left(1 + \theta^2 M_{\s}^2\right),
\end{eqnarray}
where $\phi$ is the chiral compensator, $M_{\s}$ is the scale of
supersymmetry breaking in the hidden sector, and the RHS refers to
only the scalar components. The leading tadpole divergences are
quadratic and appear at two-loop order for the first two operators in
eq.(\ref{ops}). In our case, the diagrams responsible are shown in
Figures~4a and 4b, and they lead to the terms
\begin{equation}
\frac{\lambda^\prime k}{3 (16\pi^2)^2} (\phi_N + \phi^*_N) M_{\P}
m_{3/2}^2 +
\frac{\lambda^\prime k}{3 (16\pi^2)^2} (F_N + F^*_N) M_{\P} m_{3/2}
\end{equation}
and
\begin{equation}
 \frac{\lambda^\prime\lambda}{(16\pi^2)^2}  (\phi_N + \phi^*_N) M_{\P}
m_{3/2}^2
+ \frac{\lambda^\prime\lambda}{(16\pi^2)^2}  (F_N + F^*_N) M_{\P} m_{3/2}
\end{equation}
respectively, where we have taken the cut-off to be
$\Lambda\sim~M_{\P}$ and introduced the gravitino mass
$m_{3/2}\sim\sqrt{M_{\s}^2/M_{\P}}$. Here $\lambda$ and $k$ are the
Higgs sector Yukawa couplings defined earlier.
The third term in eq.(\ref{ops})
gives rise to a divergence at three-loop order as shown in Figure~4c
and the calculation is a little more tricky. Using the perturbation
theory rules of Ref.\cite{destab}, quadratic divergences are indeed
found to arise of the form
\begin{equation}
 \frac{\lambda^\prime\lambda^2 k}{(16\pi^2)^3}
  (\phi_N + \phi^*_N) M_{\P} m_{3/2}^2
\end{equation}
and
\begin{equation}
\frac{\lambda^\prime\lambda^2 k}{(16\pi^2)^3}
  (F_N + F^*_N) M_{\P} m_{3/2},
\end{equation}
where we have replaced a quadratically divergent three loop integral
with a cut-off, $M_{\P}^2$. All of these terms naturally drive the vev
of the singlet (and hence of $H_1$, $H_2$) to the hidden sector scale,
$\langle{x}\rangle\approx\sqrt{m_{3/2}M_{\P}}\sim10^{11}\gev$. If we
wish to avoid the reappearance of the hierarchy problem, these terms
should be smaller than $\sim(\phi_N+\phi^*_N)m_{3/2}^3$ or
$\sim(F_N+F^*_N)m_{3/2}^2$. Even for the three loop diagram this
requires
\begin{equation}
\label{hierbound}
 \lambda^\prime\leqsim 3\times 10^{-11},
\end{equation}
where we have taken $m_{3/2}\sim M_W$.
Clearly this bound is only approximate, since we do not know the
precise values of the Yukawa couplings $\lambda$ and $k$, which we have taken
here to be of $\O(1)$. However, it should also be borne in mind that
one would like to have control over the scale of electroweak symmetry
breaking. That is we do not wish the mass of the $W$ to depend
strongly on the (unknown) physics at the Planck scale, i.e. on
$\lambda^\prime$. In order to achieve this, the above bound should be
tightened even further.

The bound in eq.(\ref{hierbound}) is clearly incompatible with that in
eq.(\ref{nucbound}) required for successful nucleosynthesis, and we
conclude that the NMSSM at least in the models with `minimal' K\"ahler
potentials has either a domain wall problem or a hierarchy problem.

\section{A Solution to the Hierarchy Problem}

Is it possible that we can solve these problems by allowing the hidden
and visible sectors to mix? In this section we shall see that the
answer is yes for the destabilising divergences, but no for the
naturalness problem. In other words, we are able to regain
perturbative control over the scale of electroweak symmetry breaking,
but we find, quite generally, that certain couplings must be set by
hand initially to be small.  This leads to a naturalness problem of at
least one part in $10^{9}$.

In order to eliminate destablising divergences, we must drop our
insistence on minimality in the K\"ahler potential, by allowing the
hidden and visible sectors to mix. In this case models similar to the
NMSSM can be constructed. We use a mechanism similar to that in
ref.\cite{gm}, and find that models with (standard model) singlets can
have naturally large $N^2$, $N^3$ and $\mu$ terms.

The Giudice-Masiero mechanism~\cite{gm} seeks to solve the $\mu$
problem for the MSSM by generating it via the K\"ahler potential. That
is we have
\begin{equation}
{\cal G} = y^{i} y^{\dagger}_{i}
 + z z^{\dagger} + \left(\frac{\alpha}{M_{\P}}z^{\dagger} H_{1} H_{2}
+\hc\right) +M_{\P}^2 \ln\left|\frac{f(z)+g(y)}{M_{\P}^3}\right|^2
\end{equation}
where the $y_{i}$ fields belong to the visible sector, and the $z$
singlet field belongs to the hidden sector.  $\cal{G}$ is K\"ahler
invariant. The label `hidden' is justified when we take the ``flat''
limit $M_{\P}\rightarrow\infty$ in the effective potential (keeping
$M_{\s}^2/M_{\P}$ fixed), and find that the $z$ field, which acquires
a vev of $\O(M_{\P})$, decouples from the visible sector, apart from
inducing soft SUSY breaking terms {\em and} a $\mu$ term proportional
to $\alpha$, via gravitational couplings. These are all of
$\O(M_{\s}^2/M_{\P})$, where $M_{\s}$ is the aforementioned scale of
SUSY breaking in the hidden sector which we introduce by hand.  Having
introduced a new coupling between the visible and hidden sectors, we
must invoke some symmetry which forbids other couplings as well as a
coupling $M_{\P}H_{1}H_{2}$ in the superpotential. This could be a
Peccei-Quinn symmetry, a discrete symmetry, or a gauged or global $R$
symmetry. In addition the presence of a new symmetry rules out the
simplest version of the Polonyi model (which in view of its severe
cosmological problems \cite{polonyi} might not be such a bad thing).

For the next-to-minimal choice of K\"ahler potential above, the
terms in the scalar potential are
\begin{equation}
 V_{\rm scalar} = \hat{g}_{i} \hat{g}^{i} + m_{3/2}^2 y_{i}y^{i}
                  + m^{\dagger}\left[y^{i}g_{i}+(A-3)\hat{g}^{(3)}
                  + (B-2)m_{3/2} \mu H_{1} H_{2} + \hc\right],
\end{equation}
where $\hat{g}^{(3)}$ are the trilinear terms of the superpotential,
rescaled according to
\begin{equation}
 \hat{g}^{(3)} = \langle\exp{(z z^{\dagger}/2 M_{\P}^2)}\rangle g^{(3)}.
\end{equation}
Here $\hat{g}$ is the new low energy superpotential including the
$\mu$ term
\begin{equation}
 \hat{g} = \hat{g}^{(3)} + \mu H_{1} H_{2},
\end{equation}
and $m_{3/2}$ is the gravitino mass
\begin{equation}
 m_{3/2} = \langle \exp{(z z^{\dagger} / 2 M_{\P}^2)} f^{(2)}\rangle,
\end{equation}
where the vev of $f^{(2)} = M_{\s}^2/M_{\P}$ is set by hand such
that $M_{\s}\sim10^{11}$ GeV. The $\mu$ term is given by
\begin{equation}
 |\mu| = \left|\alpha m \left\langle\frac{M_{\P}f_z}{f}\right\rangle\right|.
\end{equation}
Applying the constraint of vanishing cosmological constant, the authors of
Ref.\cite{gm} found
\begin{eqnarray}
 B &= &(2 A-3)/(A-3) \nonumber\\
 |\mu| &= &|m \alpha(A-3)/\sqrt{3}|,
\end{eqnarray}
where $A$ is the universal trilinear scalar coupling, $A=\sqrt{3}
\langle z/M_{\P}\rangle $.
Now let us apply the same mechanism to a model with MSSM singlets, $N$.
The most obvious extension is to choose the K\"ahler potential
\begin{equation}
 {\cal G} = y^{i} y^{\dagger}_{i} + z z^{\dagger}
  + \left(\frac{\alpha}{M_{\P}}z^{\dagger} H_{1} H_{2}
  + \frac{\alpha^\prime}{M_{\P}}z^{\dagger} N^2
  + \hc\right) +M_{\P}^2 \ln\left|\frac{f(z)+g(y)}{M_{\P}^3}\right|^2,
\end{equation}
where, in this case, $f(y)$ is the superpotential of the NMSSM.  The
hidden sector field, $z$, has the opposite charge to $N$ under the
$Z_{3}$ symmetry so that the full theory is $Z_{3}$-invariant.  In
this case $Z_{3}$ is broken spontaneously at the Planck scale and the
resulting domain walls are presumably removed during inflation.  The
low energy scalar potential is
\begin{equation}
 V_{\rm scalar} = \hat{g}_{i} \hat{g}^{i} + m^2 y_{i}y^{i}
  + m^{\dagger}\left[y^{i}g_{i}+(A-3)\hat{g}^{(3)}
  + (B-2) m \mu H_{1} H_{2} + (B-2)m \mu^\prime N^2 + \hc\right],
\end{equation}
where
\begin{eqnarray}
 \hat{g} &= &\hat{g}^{(3)} + \mu H_{1} H_{2} + \mu^\prime N^2 \nonumber\\
 |\mu| &= &\left|\frac{m\alpha(A-3)}{\sqrt{3}}\right| \nonumber\\
 |\mu^\prime| &= &\left|\frac{m\alpha^\prime(A-3)}{\sqrt{3}}\right|.
\end{eqnarray}
Notice that the low energy model has generally far more terms in its
low energy lagrangian than the NMSSM. The latter (and the $Z_{3}$
symmetry) is in fact recovered when we let $\alpha
=\alpha^\prime=0$; thus we can break the $Z_{3}$ symmetry by as much
or as little as we like.

Although this model has removed the problem of destabilising
divergences (it is now no longer possible to write down any of the
divergent tadpole diagrams), it does not quite solve the naturalness
problem (i.e. the presence of small couplings unprotected by any
symmetry), since there is still the coupling $zN$ which is allowed
under the $Z_{3}$ symmetry, and which no other symmetry can
forbid. These may be set to zero by hand and will stay zero by virtue
of the nonrenormalization theorem.

One might wonder if by somehow extending the K\"ahler potential it may
be possible to exclude these terms.  As we now show however, this is
not the case, and no matter how complicated we make the Lagrangian,
the naturalness problem associated with the absence of the $z N$
couplings stays with us. Consider the most general supergravity
Lagrangian, in which the only requirement we make is that the
superpotential contains the terms
\begin{equation}
 \delta g(y) = \frac{k^{abc}(\xi)}{3! 3} N_{a} N_{b} N_{c}
+ \lambda^{abc}(\xi) (H_{1}H_{2})_{ab} N_{c},
\end{equation}
where $a,b,c$ are indices representing some symmetry group (discrete
or otherwise), and the couplings are holomorphic function of the
hidden sector fields, $\xi_{a}=z_{a}/M_{\P}$. The breaking of $Z_{3}$
symmetry in the visible sector by operators of dimension-5, requires
that we also include at least one of the operators,
\begin{eqnarray}
\label{ops2}
 && \Lambda^{ab}(\xi,\bar{\xi}) N_{a} N_{b} \nonumber\\
 && \Lambda^{ab}(\xi,\bar{\xi}) (H_{1} H_{2})_{ab} \nonumber\\
 && \Lambda^{ab}_c(\xi,\bar{\xi}) N_{a} N_{b} N^{\dagger c} \nonumber\\
 && \Lambda^{ab}_c(\xi,\bar{\xi}) (H_{1} H_{2})_{ab} N^{\dagger c}
     \nonumber\\
 && \Lambda^{abcd}(\xi) N_{a} N_{b} N_{c} N_{d} \nonumber\\
 && \Lambda^{abcd}(\xi) N_{a} N_{b} (H_{1} H_{2})_{cd} \nonumber\\
 && \Lambda^{abcd}(\xi) (H_{1} H_{2})_{ab}(H_{1} H_{2})_{cd},
\end{eqnarray}
where the first four operators give dimension-5 operators if they
appear in the K\"ahler potential or superpotential, but the last three
operators must appear in the superpotential, hence their couplings are
holomorphic functions of the hidden sector fields. If we make the
assumption that the couplings $k^{abc}$ and $\lambda^{abc}$ are
invertible, then corresponding to each of the operators above, there
is an additional invariant operator which is some function of the
hidden sector fields multiplied by $N_{a}$. These are, respectively,
\begin{eqnarray}
\label{ops3}
 && \Lambda^{\dagger}_{ab} (k^{-1})^{\dagger abc} N_{c} \nonumber\\
 && \Lambda^{\dagger}_{ab} (\lambda^{-1})^{\dagger abc} N_{c} \nonumber\\
 && \Lambda^{ab}_c (k^{-1})_{abd} (k^{-1})^{\dagger cde} N_{e} \nonumber\\
 && \Lambda^{ab}_c (\lambda^{-1})_{abd} (k^{-1})^{\dagger cde} N_{e}
     \nonumber\\
 && \Lambda^{abcd}(k^{-1})_{abe}(k^{-1})_{cdf}(k^{-1})^{\dagger efg} N_{g}
     \nonumber\\
 && \Lambda^{abcd}(k^{-1})_{abe}(\lambda^{-1})_{cdf}(k^{-1})^{\dagger efg}
     N_g \nonumber\\
 && \Lambda^{abcd}(\lambda^{-1})_{abe}(\lambda^{-1})_{cdf}
     (k^{-1})^{\dagger efg} N_{g}.
\end{eqnarray}
The least damage to the effective potential occurs if these terms
appear in the K\"ahler potential, in which case we find terms of the
form
\begin{equation}
 m_{3/2}^2 M_{\P} \phi_N + \hc
\end{equation}
appearing in the effective potential. Thus the natural scale of the
singlet vev is $\sim10^{11}$ GeV and since it should be less than the
electroweak scale, this constitutes a naturalness problem of at least
one part in $10^{9}$.

\section{Conclusions}

Before we conclude there are a few escape clauses which
should be mentioned none of which however are very appealing:

\begin{enumerate}

\item The most obvious is to introduce the $\mu$ term into the
superpotential by the mechanism of Ref.~\cite{gm} and simply set to
zero all of the operators which might give $N$ a large vev. (Although
this appears to be rather unaesthetic, one might remark that the
naturalness problem which results is no worse than that already with
us due to the smallness of $M_{\s}$ compared to the Planck mass. Since
the ``unnaturalness'' is of the same order, it may even be possible to
construct the K\"ahler potential so that the two naturalness problems
are connected.)

\item Alternatively one can invoke inflation at the weak scale to
remove all the domain walls, just as has been suggested in the context
of other unwanted relics, e.g. string moduli~\cite{weakinfl}. However
such a scenario must be very finely tuned --- the domain walls must be
adequately diluted without erasing the density perturbations generated
by inflation at the GUT scale~\cite{gutinfl}). (Although density
perturbations are also generated during weak scale inflation, the
small value of the Hubble parameter would make these too small to
account for the microwave background anisotropies observed by COBE.)
Secondly, the reheat temperature must be high enough for both
successful baryogenesis and nucleosynthesis. We are not aware of any
compelling candidate for the required scalar field.

\item The $Z_3$ symmetry could be broken at a high scale, $M_{\con}$,
in the visible sector and explicit $Z_3$ breaking terms induced. This
is similar to the solution to the hierarchy problem discussed earlier,
with the ``advantage'' that the fine tuning is driven by $M_{\con}$
rather than $M_{\P}$. However this will still entail a fine tuning of
approximately one part in $10^{12}$, since in order for the walls to
be inflated away, $M_{\con}$ should exceed the inflationary scale of
$\sim10^{14}\gev$ as deduced from normalisation to the COBE
data~\cite{gutinfl}. Otherwise one would have to invoke a second epoch
of inflation at an intermediate scale, with its own attendant problems
(see above).

\item The $Z_3$ symmetry could be made anomalous by adding extra
fields to the theory which couple to $SU(3)_{\c}$ (for example an
additional generation). In this case the symmetry is broken
non-perturbatively at the quark-hadron phase transition, and the walls
collapse very soon thereafter~\cite{ptww}.  However, it is difficult
to see how this constitutes a solution to fine-tuning, since at the
same time it seems to preclude a solution to the strong CP problem as
discussed in Ref.\cite{ptww}.

\item The $Z_3$ symmetry could be embedded in a continuous gauge or
global group which is broken at some high scale. This is the
Lazarides-Shafi mechanism \cite{ls}, in which the apparent discrete
symmetry is a subgroup of the centre of the continuous group. In this
case only $U(1)$, $SU(3n)$ (where $n$ is an integer) and $E_6$ are
suitable candidates (see for example Ref.\cite{kim}). After the
electroweak phase transition, one expects only a network of walls
bounded by strings to form and then quickly collapse \cite{ls}.

\end{enumerate}

To summarize, we have shown that the domain wall problem in the NMSSM
causes it to be ruled out on cosmological grounds unless we break the
$Z_{3}$ symmetry of the model explicitly. The breaking may be driven
by terms which are non-renormalisable and have no direct effect on the
low energy theory. However their introduction will in general generate
terms which destabilise the hierarchy.  In models with ``minimal''
K\"ahler potentials, we have shown that there are no
non-renormalisable operators which can be added to the superpotential
with a coefficient which is simultaneously large enough to solve the
cosmological problem and small enough to avoid reintroducing the
hierarchy problem.  Furthermore, if any of these operators are allowed
by the symmetries of the theory at the supergravity scale, then there
is no possible symmetry which could prevent the existence of an
operator $zN$ in the superpotential whose coefficient must be
$\leqsim10^{-17}$. If we allow mixing between the hidden and visible
sectors, the reintroduction of the hierarchy problem can be avoided,
and the naturalness problem can be formulated in a way very similar to
the $\mu$ problem in the MSSM.  However, even here we must arbitrarily
select coefficients of certain dangerous operators to be of
$\O(10^{-9})$ or less once we have aranged for a $\mu$ parameter of a
reasonable size, and we have also reintroduced the $\mu$ problem which
the model was at least partly designed to solve. Thus we conclude that
the parameters in the NMSSM must be very strongly fine tuned if we
wish to avoid both the cosmological problems associated with domain
walls and the hierarchy problem, and hence that the model suffers from
severe naturalness problems.

\vspace{1cm}
\noindent
{\bf \Large Acknowledgements:} We would like to thank Graham Ross for
encouraging us to undertake this study and for many discussions. We
are also grateful to H.~Dreiner, U.~Ellwanger, E.W.~Kolb and M.~Rausch
de Traubenberg for their criticism and comments, and especially to
J.~Bagger for discussions concerning destabilising divergences.

\newpage

{\center {\bf Figure Captions}}

\noindent
{\bf Figure 1a,b}\\
An example of a wall configuration with the singlet vev $x=10\nu$. Here
we have chosen $\tan\beta=2$, $\lambda=k=0.2$, $A_k=A_\lambda=200$GeV.
Total surface energy density is $8.6\times 10^8\hbox{\rm GeV}^3$.
Figure 1a shows the values of the three scalar fields as a function of
position; Figure 1b shows the energy density in the wall relative to
the vacuum.

\noindent
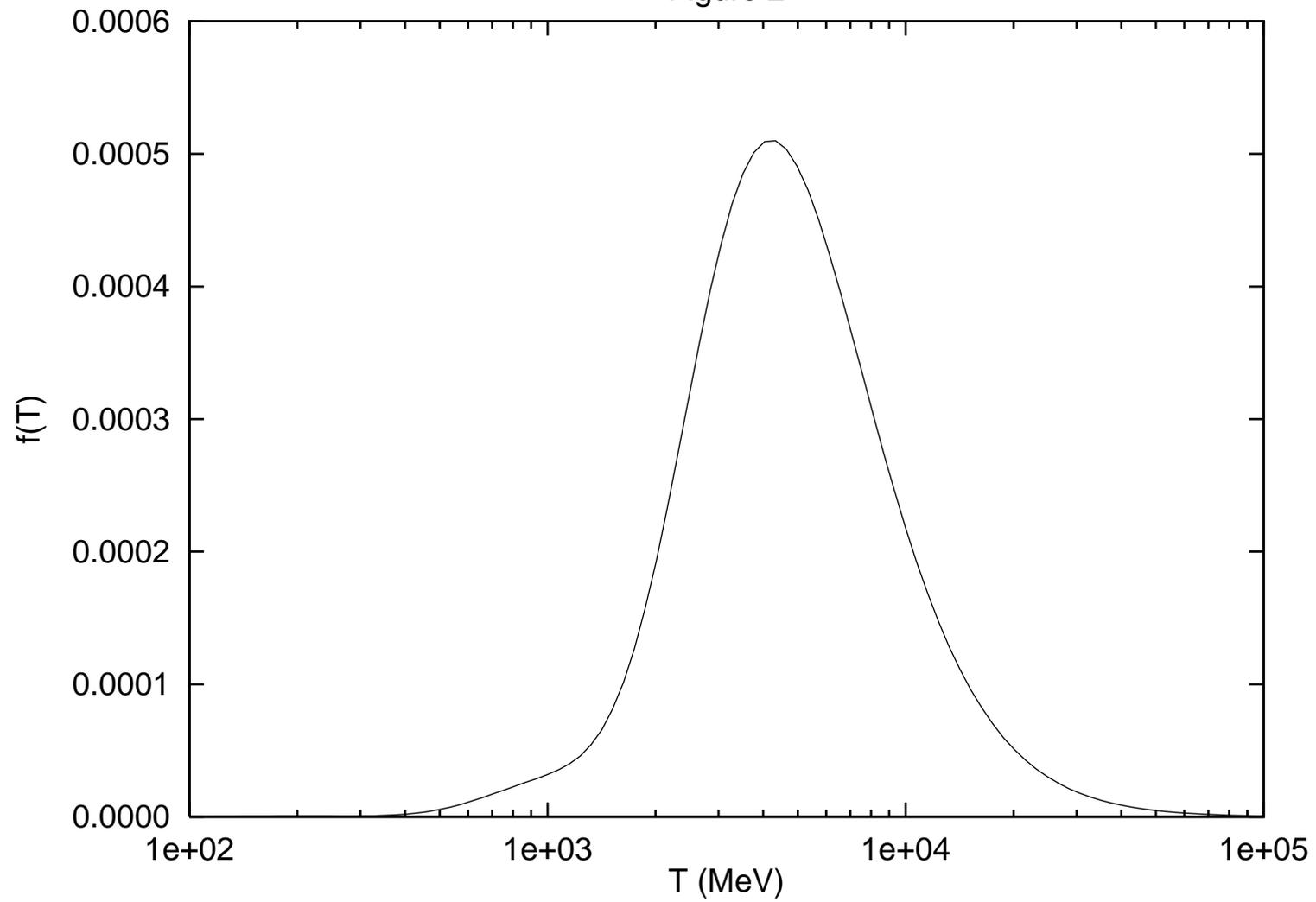
{\bf Figure 2}\\
The function $f(T)$ (see eq.\ref{friction}) related to friction plotted
against temperature, as discussed in the text.

\noindent
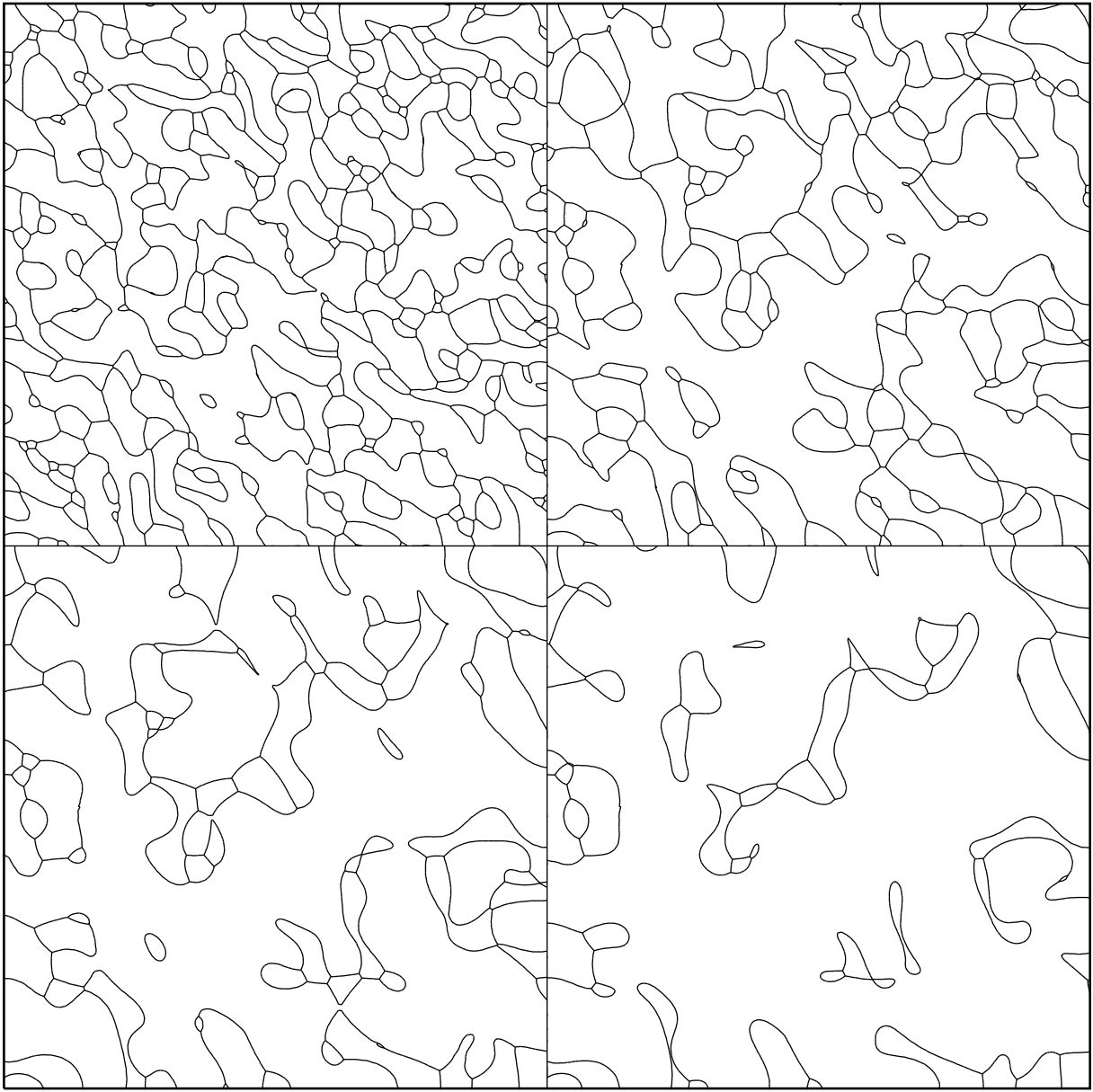
{\bf Figure 3}\\
A typical example of the evolution of the wall network with pressure.
The figure shows the wall network at four times separated by intervals
of $10^{-10}$sec, with a pressure of term of order $\sigma
M_W^2/M_{\P}$, beginning at the time when pressure starts to dominate
the evolution.

\noindent
{\bf Figure 4a,b,c}\\
The three dangerous diagrams for each of the three operators which
can destabilise the hierarchy.

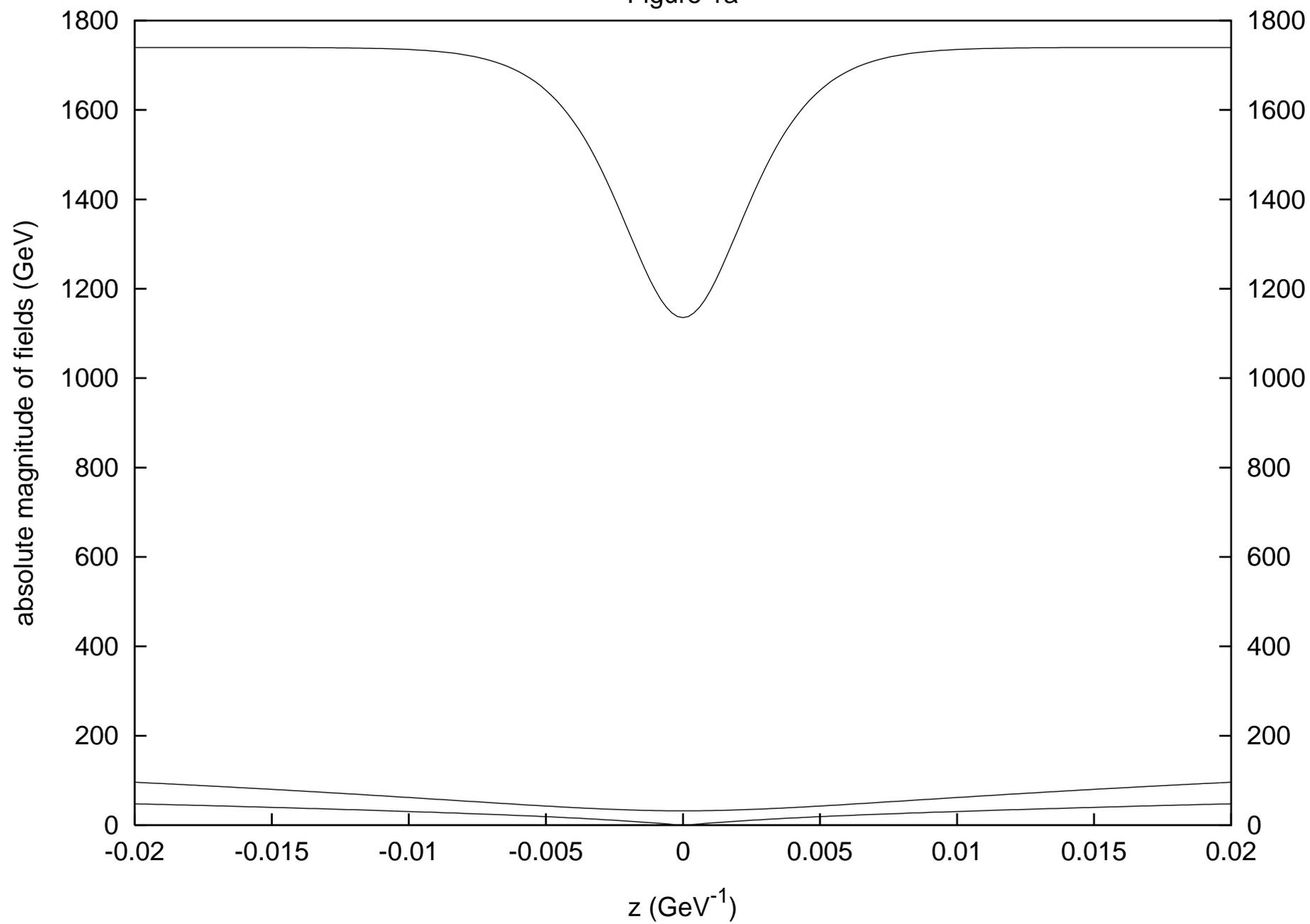
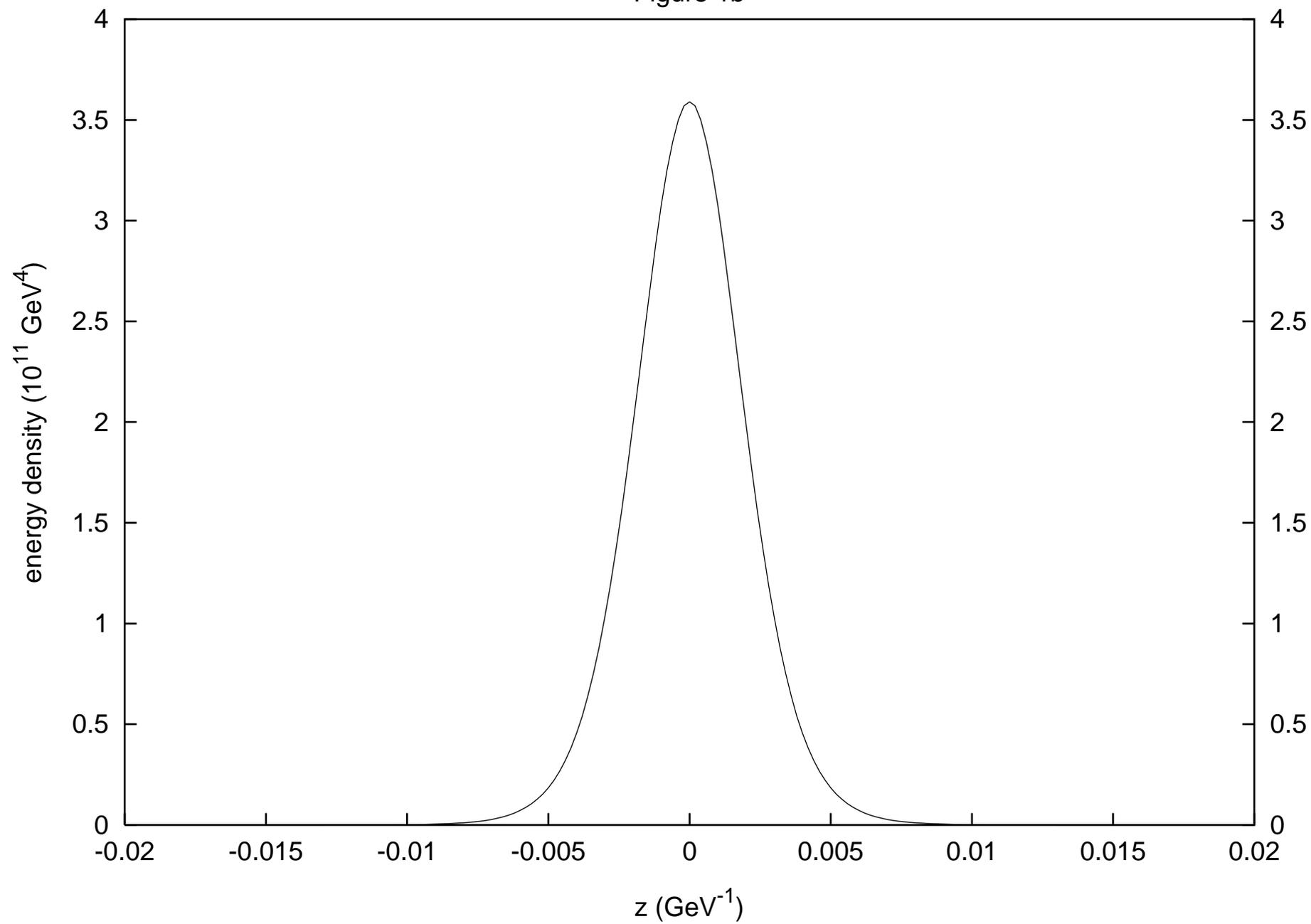
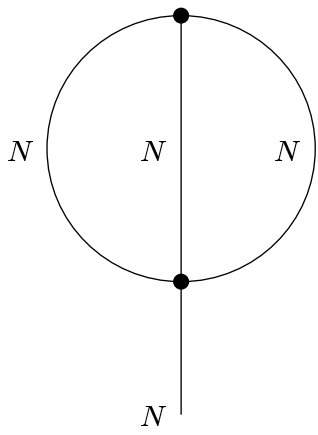
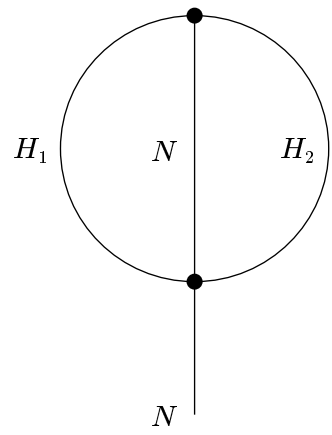
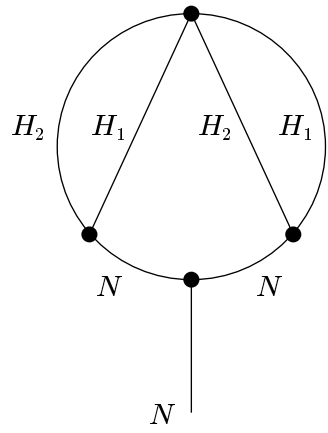
\end{document}